\documentclass[conference]{IEEEtran}
\IEEEoverridecommandlockouts

\usepackage{cite}
\usepackage{amsmath,amssymb,amsfonts}
\usepackage{algorithm}
\usepackage{algpseudocode}
\usepackage{graphicx}
\usepackage{textcomp}
\usepackage{multirow}
\usepackage{csquotes}
\usepackage{hyperref}
\usepackage{listings}
\usepackage{lettrine}
\usepackage{array}
\usepackage{stfloats}
\usepackage{url}
\usepackage{verbatim}
\usepackage{academicons}
\usepackage{upgreek}
\usepackage{siunitx}
\usepackage{adjustbox}
\usepackage{rotating}
\usepackage[inline]{enumitem}
\usepackage{relsize}
\usepackage{longtable}
\usepackage{xcolor}
\usepackage{threeparttable}
\usepackage[caption=false]{subfig}
\usepackage{tabu}
\usepackage{color, colortbl}
\usepackage{booktabs}
 
\usepackage{pifont}
\usepackage{tabularx}

\begin{document}

\title{Accelerating Sensor Fusion in Neuromorphic Computing: A Case Study on Loihi-2}

\makeatletter
\newcommand{\linebreakand}{%
  \end{@IEEEauthorhalign}
  \hfill\mbox{}\par
  \mbox{}\hfill\begin{@IEEEauthorhalign}
}
\makeatother

\author{%
\IEEEauthorblockN{Murat Isik\IEEEauthorrefmark{1}}
\IEEEauthorblockA{%
\textit{Drexel University}\\
Philadelphia, USA \\
Email: mci38@drexel.edu\\
\textit{\IEEEauthorrefmark{1}Corresponding author}}
\and

\IEEEauthorblockN{Karn Tiwari}
\IEEEauthorblockA{%
\textit{Indian Institute of Science, Bangalore}\\
Bengaluru, India \\
karntiwari@iisc.ac.in}

\and
\IEEEauthorblockN{Muhammed Burak Eryilmaz}
\IEEEauthorblockA{%
\textit{Bilkent University}\\
Ankara, Turkey \\
burak.eryilmaz@bilkent.edu.tr}

\linebreakand 
\IEEEauthorblockN{I. Can Dikmen}
\IEEEauthorblockA{%
\textit{Temsa Research \& Development Center}\\
Adana, Turkey \\
can.dikmen@temsa.com}

}


\maketitle

\vspace{-10pt}

\begin{abstract}
In our study, we utilized Intel's Loihi-2 neuromorphic chip to enhance sensor fusion in fields like robotics and autonomous systems, focusing on datasets such as AIODrive, Oxford Radar RobotCar, D-Behavior (D-Set), nuScenes by Motional, and Comma2k19. Our research demonstrated that Loihi-2, using spiking neural networks, significantly outperformed traditional computing methods in speed and energy efficiency. Compared to conventional CPUs and GPUs, Loihi-2 showed remarkable energy efficiency, being over 100 times more efficient than a CPU and nearly 30 times more than a GPU. Additionally, our Loihi-2 implementation achieved faster processing speeds on various datasets, marking a substantial advancement over existing state-of-the-art implementations. This paper also discusses the specific challenges encountered during the implementation and optimization processes, providing insights into the architectural innovations of Loihi-2 that contribute to its superior performance.
\end{abstract}


\section{Introduction}

The rapid advancement of computational science and engineering has perpetually driven the quest for enhanced efficiency and processing speed, particularly in sensor fusion. Sensor fusion, a pivotal component in cutting-edge applications such as autonomous vehicles, robotics, and sophisticated monitoring systems, epitomizes the need to process complex data from diverse sources efficiently. Traditional computational paradigms, while effective, grapple with challenges in scalability, speed, and energy efficiency when it comes to the multifaceted nature of sensor fusion. This integration of sensory data from multiple sources is crucial for creating a more comprehensive understanding of the environment, which is paramount in modern technological applications, enabling enhanced perception and decision-making capabilities in autonomous systems \cite{elmenreich2002introduction, yeong2021sensor, yu2023automated, yao2023radar}. However, the traditional computational methods for sensor fusion are increasingly strained by the demands for real-time processing, accuracy, and the handling of large volumes of diverse data. These challenges necessitate a paradigm shift to more capable and efficient computing methods. In this context, neuromorphic computing emerges as a transformative solution. Inspired by the neural structure and functioning of the human brain, neuromorphic computing offers a novel approach to data processing. Its ability to mimic biological neural networks promises significant improvements in speed and energy efficiency, crucial for real-time and complex tasks like sensor fusion \cite{cho2022progress, zou2022eventhd, isik2023hpcneuronet, inadagbo2023exploiting}. Intel’s Loihi-2, a leading-edge neuromorphic chip, exemplifies this technological leap. Characterized by its spiking neural networks (SNNs), Loihi-2 offers an innovative architecture tailored to the complexities of sensor fusion tasks. This study focuses on Loihi-2, exploring its capabilities in enhancing the efficiency and speed of sensor fusion processes. The potential benefits of employing Loihi-2 for sensor fusion are manifold and hypothesized to include accelerated processing speed, heightened energy efficiency, and improved adaptability to diverse sensor modalities and dynamic environments. However, there is a gap in the current research literature, particularly in applying neuromorphic computing like Loihi-2 in sensor fusion tasks. This paper aims to bridge this gap by providing empirical insights into the performance and advantages of Loihi-2 in sensor fusion applications \cite{harbour2023real, orchard2021efficient, marchisio2023embedded}. Our research objectives are twofold: firstly, to demonstrate the efficacy of Loihi-2 in accelerating sensor fusion processes, and secondly, to contribute novel findings to the field of neuromorphic computing, underlining its practical applications and potential in various technological domains. Through this study, we aim to set a precedent for future research and development in neuromorphic computing, particularly in its application to sensor fusion and related fields. The exploration of sensor fusion through the lens of neuromorphic computing, specifically via the capabilities of Loihi-2, presents an opportunity to address the limitations of traditional computational methods. It opens a pathway to more efficient, accurate, and real-time processing of sensory data, which is essential in the rapidly evolving landscape of technology and automation. We propose a design methodology for sensor fusion applications consisting of three components.

\begin{itemize}
\item  Our research introduces a cutting-edge approach for integrating data from various sensors like visual, auditory, and inertial, using the Loihi-2 chip.
\item We conducted comprehensive evaluations using multiple datasets (AIODrive, Oxford Radar RobotCar, etc.) to demonstrate the effectiveness of our methods in different real-world scenarios.
\item  Our study provides a detailed comparison of the Loihi-2 chip's performance against traditional CPUs and GPUs, showcasing its superior energy efficiency and processing speed.
\end{itemize}

Each component of our approach contributes to the enhanced performance of sensor fusion tasks and exemplifies the potential of neuromorphic computing to revolutionize real-time data processing in various applications.

\section{Background}

This section provides a comprehensive overview of the fundamental concepts essential to our study and its related work. It encompasses the domains of sensor fusion, neuromorphic computing, and the Intel Loihi-2 chip.

\subsection{Sensor Fusion}

Sensor fusion involves integrating data information from multiple sensors to produce more accurate, reliable, and contextual information than possible from any single sensor source. This technology is fundamental in applications ranging from autonomous vehicles and robotics to healthcare and environmental monitoring \cite{sasiadek2002sensor, yeong2021sensor}.

Various algorithms drive sensor fusion processes, aiming to merge data streams intelligently. These algorithms include rule-based systems, statistical methods such as Kalman filters, and machine learning techniques like neural networks. Rule-based systems establish decision-making guidelines based on predefined rules, while statistical methods focus on optimizing estimates by considering both the sensor measurements and their uncertainties. Machine learning algorithms enable systems to learn and adapt to dynamic environments, providing a more flexible approach to sensor fusion \cite{elmenreich2002introduction, amaro2016survey, dong2009advances}.

Sensor fusion often suffers from the diversity and heterogeneity of sensor data as different sensors generate information in different formats and rates. Integrating data from different sensors requires sophisticated algorithms and computational approaches to ensure accuracy and reliability. Additionally, the real-time nature of many applications demands efficient data processing and the ability to handle dynamic and unpredictable environments. Calibration discrepancies, sensor noise, and uncertainties compound the complexity of sensor fusion tasks, necessitating robust methods to filter and interpret data accurately. The sheer volume of data generated by modern sensors poses scalability challenges, straining from traditional computing architectures. Overcoming these challenges is crucial for unleashing sensor fusion's full potential in enhancing technological systems' capabilities \cite{sasiadek2002sensor, yeong2021sensor, van2020multi}.

\subsection{Neuromorphic Computing}

This subsection explores the foundational principles of neuromorphic computing, tracing its evolution and examining its transformative potential in addressing the limitations inherent in traditional computing methods. Neuromorphic computing is inspired by the architecture and functionality of the human brain, leveraging parallel processing, adaptive learning, and efficient energy utilization. Unlike traditional computing, which often relies on sequential processing, neuromorphic systems excel in handling parallel information, making them particularly well-suited for data-intensive tasks like sensor fusion \cite{furber2016large, huynh2022implementing, isik2023survey, isik2024neurosec, isik2024advancing}. The evolution of neuromorphic computing has seen advancements in hardware architectures, such as Intel's Loihi processor, designed to emulate the brain's neural networks. These systems offer a departure from conventional binary logic, introducing spiking neural networks and event-driven computation to mimic the brain's processing mechanisms better \cite{davies2021advancing, orchard2021efficient}. The potential of neuromorphic computing lies in its ability to overcome the limitations of traditional computing methods, especially in scenarios demanding real-time processing and adaptability. In data-intensive tasks like sensor fusion, where diverse information streams must be seamlessly integrated, neuromorphic systems showcase advantages in parallelism, efficiency, and the ability to learn from experience \cite{roy2019towards, yu2021multi}. By mimicking the brain's sophisticated processing, neuromorphic computing holds promise for revolutionizing sensor fusion applications. It can enhance the speed, accuracy, and adaptability of processing sensor data, leading to more robust and intelligent decision-making systems. As research and development in neuromorphic computing continue progressing, its potential to redefine the landscape of computational methodologies, particularly in data-intensive tasks like sensor fusion, becomes increasingly evident \cite{aimone2022review}.

\subsection{Intel’s Loihi-2 Neuromorphic Chip}

Loihi-2, the latest iteration in Intel Labs' series of advanced neuromorphic chips, represents a significant leap forward from its predecessor, Loihi. Released in 2022 as a cutting-edge neuromorphic research test chip, Loihi-2 offers enhanced scalability, processing speed, and energy efficiency. Fabricated on an Intel 4 process, Loihi-2 boasts 128 neuromorphic cores and is classified as a multi-core IC. It incorporates an asynchronous SNN for adaptive, self-modifying, and event-driven parallel computations, optimizing efficiency for learning and inference processes. The inclusion of a programmable microcode learning engine facilitates on-chip SNN training. The accompanying board, Kapoho Point, exemplifies Loihi-2's prowess with its 2-stacked configuration housing eight chips in a 4x4 arrangement, resulting in an impressive ensemble of 1,024 neuromorphic cores, 960,000,000 synapses, and 8.4 million neurons \cite{davies2021advancing, orchard2021efficient}. Loihi-2's architecture, with its SNNs, offers a unique and efficient approach to processing information, mimicking the parallelism and adaptability of the human brain. This intrinsic capability aligns seamlessly with the demands of sensor fusion, where real-time integration of diverse data sources is paramount. The asynchronous and event-driven nature of Loihi-2's SNNs enhances its suitability for handling sensor data's dynamic and time-sensitive nature. Moreover, the chip's increased scalability and processing speed empower it to manage the complexities of merging information from multiple sensors in real-time. As we delve into the intricacies of Loihi-2's architecture, it becomes evident that its design nuances make it an ideal candidate for advancing the field of sensor fusion, offering a promising avenue for more efficient and intelligent information integration \cite{davies2021advancing, roy2019towards, pawlak2024exploring}.

\begin{figure}[h]
\centering
\includegraphics[width = 0.5\textwidth]{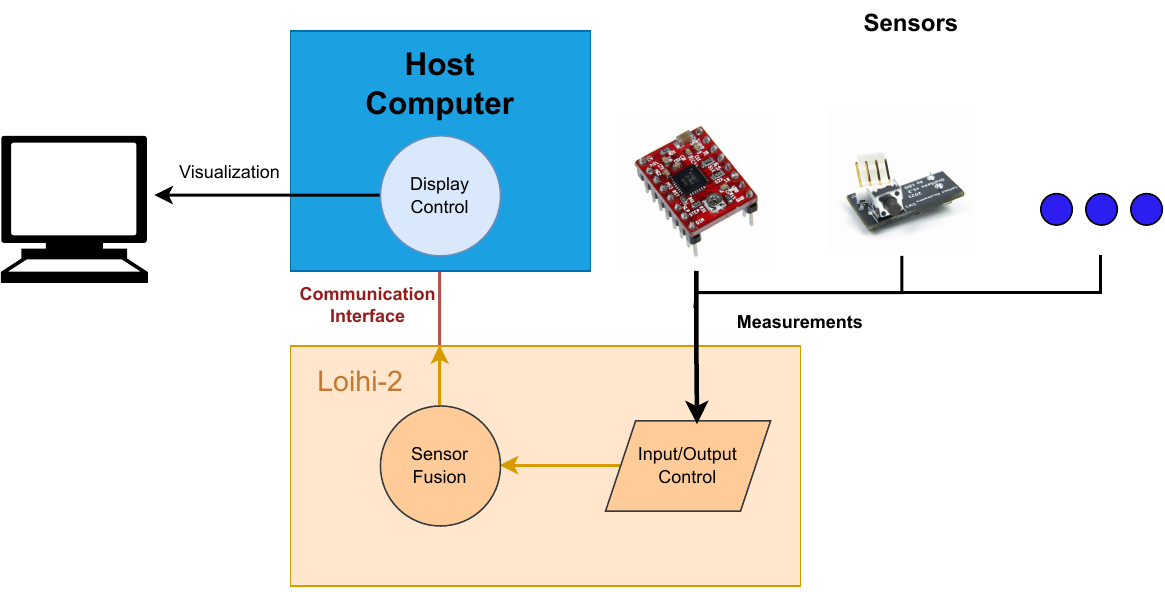}
\caption{System architecture integrating the Loihi-2 neuromorphic chip for advanced sensor fusion.}
\label{fig1}
\end{figure}

\autoref{fig1} presents a simplified computing system diagram incorporating the Intel Loihi-2 chip, designed to combine data from different sensors. The main computer interface allows users to see visual representations of data and control what is displayed. This part of the system effectively manages how data is shown to users. The Loihi-2 chip is central to the system's architecture, showcasing the system's ability to process information. This chip has a 'Sensor Fusion' capability, which means it can take in and integrate data from various sensors. It also manages the incoming and outgoing data, a process essential for controlling the flow of information. On the diagram's right side, we see various sensors acting as the system's starting points for data collection. They are shown as capturing and sending out information, likely in spikes, a specialized signal used in systems like Loihi-2. The 'Communication Interface' connects the main computer to the Loihi-2 chip. It represents the system's method for exchanging data, possibly through a standard communication protocol like Ethernet, facilitating a smooth and integrated data flow.

\section{Methods}

This section outlines the methodology employed in our study to explore the efficacy of Intel's Loihi-2 neuromorphic chip in accelerating sensor fusion.

\subsection{Dataset}

The data used in our experiments are carefully selected from renowned datasets in autonomous driving and sensor fusion, reflecting a range of real-world scenarios and sensor modalities. These datasets are complemented by synthetic data generated to test specific conditions and hypotheses. The datasets include:

\begin{itemize}
\item \textbf{AIODrive Dataset}: Offers a comprehensive collection of multimodal sensory data, including high-resolution camera images, LIDAR point clouds, GPS trajectories, and IMU data across diverse environments such as urban, suburban, and highways \cite{weng2020all}.

\item \textbf{Oxford Radar RobotCar Dataset}: Specializes in radar data complemented by LIDAR, camera, and GPS, crucial for autonomous navigation research, especially in adverse weather conditions and for radar-LIDAR fusion algorithms \cite{maddern20171, barnes2020oxford}.

\item \textbf{D-Behavior Dataset (D-Set)}: Provides data on driver behavior and traffic interaction, including cameras, radars, and CAN-Bus information, valuable for analyzing driving behaviors and developing advanced driver-assistance systems \cite{chen2018lidar}.

\item \textbf{nuScenes Dataset by Motional}: The nuScenes dataset is a comprehensive resource for autonomous driving research developed by Motional. It provides a rich collection of multimodal sensor data, including 3D bounding boxes, camera images, RADAR, and LiDAR point clouds. The dataset covers many urban driving scenarios and is particularly valuable for object detection, prediction, and segmentation tasks in complex urban environments \cite{caesar2020nuscenes, pham20203d}. The diverse and detailed nature of the dataset makes it highly suitable for developing and testing advanced algorithms in perception, sensor fusion, and autonomous navigation.

\item \textbf{Comma2k19 Dataset}: Features extensive real-world driving data, primarily from highway scenarios, including high-definition video, CAN data, and GPS, suitable for developing cruise control algorithms and lane-keeping systems \cite{schafer2018commute}.
\end{itemize}

Each dataset provides a unique perspective and data for advancing research in autonomous driving and sensor fusion, with specific focuses ranging from environmental perception to driver behavior analysis. Given their complex, multimodal nature, they are particularly well-suited for testing and developing algorithms for neuromorphic hardware like Intel's Loihi-2.

\begin{figure}[h]
\centering
\includegraphics[width = 0.3\textwidth]{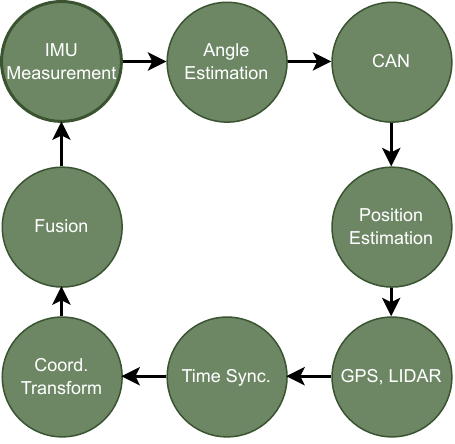}
\caption{Sensory data from cameras, LIDAR, RADAR, GPS, and IMU, processed through a neuromorphic computing unit, exemplified for advanced autonomous driving applications.}
\label{fig2}
\end{figure}

\autoref{fig2} emphasizes the integration and processing of data from multiple sensors through a neuromorphic computing unit, which is a type of computing architecture inspired by the human brain, known for its efficiency in processing complex data sets, a key feature in autonomous driving systems. It underscores the integration of these diverse sensory inputs through advanced computing techniques, which is critical for the sophisticated task of autonomous driving.

\subsection{Overview of the Proposed Framework}

We employ SNNs, optimized for Loihi-2, for data processing. We have successfully implemented five distinct SNN models, each tailored to a specific dataset. These implementations were initially executed on conventional CPU/GPU architectures to validate their performance and functional integrity. Our primary focus has been optimizing these models for efficient computation while ensuring their compatibility with neuromorphic hardware, specifically Intel's Loihi-2 chip. The current phase of our research involves the critical step of generating an HDF5 file, a format necessary for interfacing with the Loihi-2 programming environment. This file encapsulates the trained SNN models' parameters and architecture in a format compatible with the neuromorphic hardware. Generating this file is a nuanced process, demanding careful consideration of the specific requirements and constraints of the Loihi-2 platform. Further, we are in the process of scripting the implementation for Loihi-2. This script is crucial for setting up the computational environment on the Loihi-2 chip, particularly for executing a model trained on the CIFAR-10 dataset. A significant challenge in this phase is the discrepancy in the data loader's output/input formats when interfacing with the Loihi-2 architecture. This challenge underscores the complexities of adapting neural network models from conventional computing paradigms to neuromorphic platforms. To address these challenges, our approach involves leveraging the Lava-DL library (including Slayer and Bootstrap libraries) specifically designed to develop and train SNNs for neuromorphic hardware like Loihi-2. Lava-DL offers a conducive environment for training SNNs and facilitating their transition to neuromorphic platforms. Our implementation script includes critical components for interfacing with the Loihi-2 chip. These components include defining input/output ports, ensuring the seamless exportation of the HDF5 file, and leveraging various Lava libraries such as lava-dl netx, lava.magma.core, lava.process, and lava.util. These libraries provide a rich set of tools and functionalities essential for tailoring SNNs to the unique computational architecture and processing capabilities of the Loihi-2. Adapting SNNs for execution on Intel's Loihi-2 neuromorphic chip represents a significant advancement in neuromorphic computing. It highlights the potential of neuromorphic hardware in handling complex computational tasks, particularly those involving real-time data processing and energy-efficient computation. The successful implementation of these models on Loihi-2 would validate the versatility and robustness of the SNN architectures we have developed and underscore the practical applications and benefits of neuromorphic computing in various domains.

\begin{figure}[h]
\centering
\includegraphics[width = 0.4\textwidth]{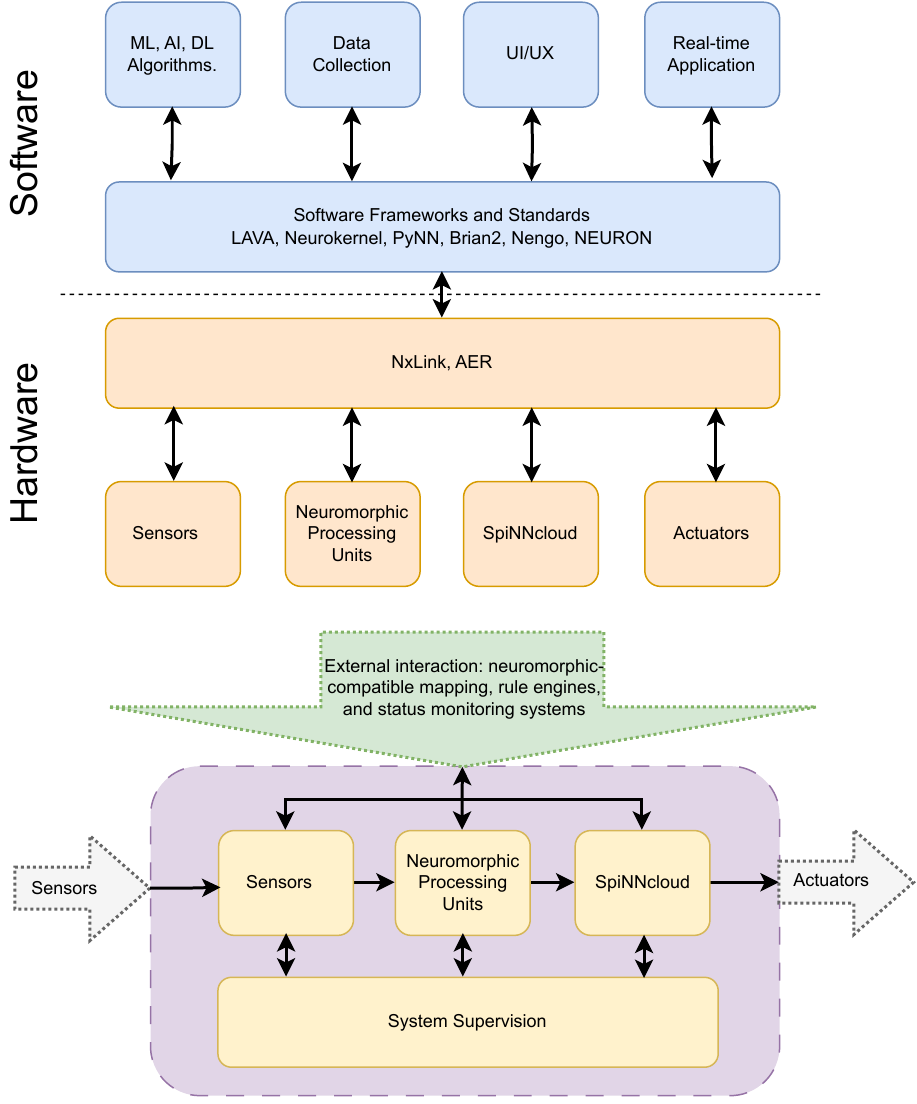}
\caption{Overview of a Neuromorphic Computing System Architecture, integrating specialized software framework. The system employs advanced communication protocols and hardware components designed for real-time data processing and interaction with the environment.}
\label{fig3}
\end{figure}

\autoref{fig3} illustrates an integrated neuromorphic computing system designed for advanced machine learning, artificial intelligence, and deep learning applications. At the software level, it employs a suite of specialized frameworks and standards such as LAVA, Neurokernel, PyNN, Brian2, Nengo, and NEURON, which are pivotal for the development and execution of neuromorphic algorithms. These tools facilitate the algorithmic translation necessary for neuromorphic computing and manage data collection, user interaction, and real-time application deployment. The system leverages communication protocols like NxLink and AER, tailored to the neuromorphic paradigm, particularly handling spike-based data transmission. At the hardware core, sensors gather real-world data, which is then processed by neuromorphic units that execute brain-inspired computational models. Actuators in the loop respond to the computational directives, and platforms such as SpiNNcloud suggest an infrastructure for distributed neural network simulations. The architecture also incorporates external interfaces for neuromorphic-compatible mapping, rule engines, and systems for status monitoring, ensuring the system’s adaptability and responsiveness to dynamic environmental conditions.

\begin{figure*}
\center \includegraphics[width=0.65\textwidth]{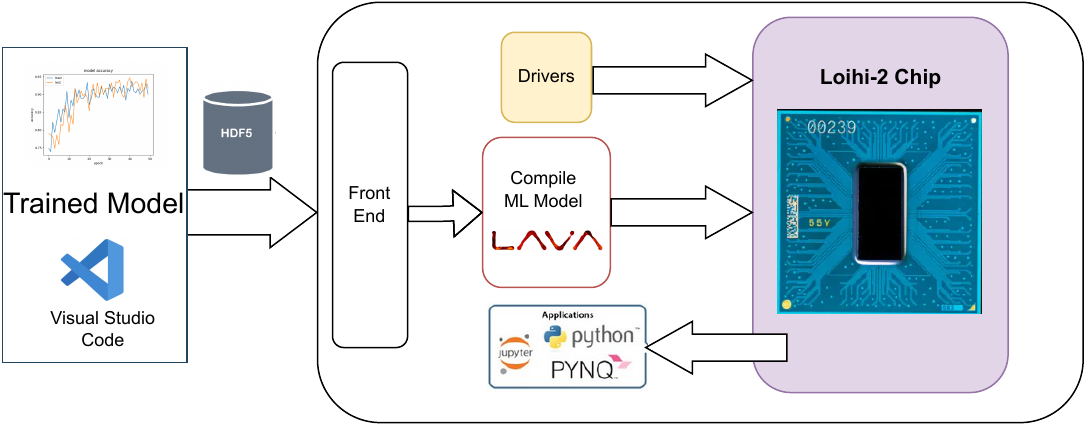}
\caption{Block Diagram of Implementation}
\label{block}
\end{figure*}

The diagram in \autoref{block} outlines implementing a machine learning model on the Loihi-2 chip, specialized in neuromorphic computing. This sequence begins with the trained model in a development environment and moves through stages of preparation and compatibility adjustments via a front-end interface. The model is then compiled into an executable form for the Loihi-2 chip using dedicated drivers and tools. Finally, the model is operational on the neuromorphic hardware, ready to perform tasks with efficiency reminiscent of biological neural systems, leveraging the unique capabilities of neuromorphic computing to process data in a brain-inspired manner.

\subsection{Additional Considerations:}
\begin{itemize}
\item \textbf{Sparsity:} Leverage the inherent sparsity of neuromorphic computing in our model to reduce computational load and increase efficiency.
\item \textbf{Neuron Model Simplification:} Simplify the neuron models to fit the dynamics of the Loihi-2 chip, which might involve using fewer synaptic operations or adjusting the neural dynamics for the hardware.
\item \textbf{Software-Hardware Compatibility:} Ensure that our model is compatible with Loihi-2's software stack, which includes NxSDK and other Loihi-specific programming tools.
\item \textbf{Neuromorphic Debugging Tools:} Utilize neuromorphic-specific debugging tools provided by Intel for the Loihi-2 chip to troubleshoot and optimize the deployment of the model.
\end{itemize}

\section{Evaluation}

The experiments are designed to evaluate the performance of sensor fusion using Loihi-2 in various aspects such as speed and efficiency, power and energy consumption.

\begin{table*}[h]
\centering
\caption{Specifications and Evaluation Results of Hardware Devices}
\label{table1}
\begin{adjustbox}{max width=\textwidth,center}
\begin{threeparttable}
\begin{tabular}{lccc}
\toprule
\textbf{Specification} & \textbf{Intel i9 12900H (CPU)} & \textbf{NVIDIA RTX 3060 (GPU)} & \textbf{Intel Loihi 2 (ASIC)} \\
\midrule
\textbf{Model} & Core i9-12900HA & GeForce RTX 3060 & Loihi 2 \\
\textbf{Technology Node} & 10nm & 8nm & 7nm \\
\textbf{Core Count} & 14 & 3,584 & 128 \\
\textbf{Precision} & Float32 & Float32 & Fixed32 \\
\textbf{Operation Frequency (MHz)} & 3700 & 1320 & 1000 \\
\textbf{Memory Technology} & DDR5 & GDDR6 & On-chip \\
\textbf{Maximum Power (W)} & 157 & 170 & Low Power \\
\textbf{\# of MAC (MOP)} & 960 & 960 & 960 \\
\textbf{Throughput (GOP/s)} & 8.67 & 304.76 & 161.11 \\
\textbf{Power (Watt)} & 28 & 80 & 1.55 \\
\textbf{Power Efficiency (GOP/s/W)} & 0.30 & 3.80 & 103.94 \\
\bottomrule
\end{tabular}
\begin{tablenotes}[flushleft]
\item Oxford Radar RobotCar dataset was implemented on these models.
\end{tablenotes}
\end{threeparttable}
\end{adjustbox}
\end{table*}

\autoref{table1} presents a detailed comparison across several crucial hardware specifications and performance metrics, including technology node, core count, precision, operational frequency, memory technology, power consumption, and other critical parameters like throughput and power efficiency. This comparison elucidates the distinctive features and advantages of the Intel i9 12900H CPU, NVIDIA RTX 3060 GPU, and Intel Loihi 2 ASIC. A notable aspect of this comparison is the power efficiency and throughput metrics, particularly for the Intel Loihi 2 ASIC. The Loihi 2 exhibits an exceptional power efficiency of 103.94 GOP/s/W, significantly outperforming the CPU and GPU. This is accompanied by a modest power requirement of only 1.55 Watts, a fraction of the power consumed by the other two devices. Despite its lower operational frequency of 1000 MHz and a relatively smaller core count of 128, the Loihi 2 ASIC demonstrates a throughput of 161.11 GOP/s, highlighting the specialized capabilities of ASICs in efficiently handling specific computational tasks. In contrast, the GPU shows a higher throughput of 304.76 GOP/s, benefitting from its large core count of 3,584 and a higher operational frequency of 1320 MHz. However, this comes at the cost of increased power consumption, evident from its 80-watt power requirement. With its balanced core count and precision, the CPU offers moderate performance in both throughput and power efficiency, suitable for a wide range of general-purpose computing tasks.

\begin{table*}
\renewcommand{\arraystretch}{1}
\setlength{\tabcolsep}{1.5pt}
\centering
\caption{Performance Metrics of Loihi-2 on Various Datasets}
\label{table2}
\small
\begin{tabular}{llllllll}
\toprule
 & \textbf{Power (W)} & \textbf{Inference Time (s)} & \textbf{Throughput (GOP/s)} & \textbf{Efficiency (pJ/OP)} & \textbf{Complexity} & \textbf{Neurons} & \textbf{Synapses} \\
\midrule
AIODrive & 2.50 & 0.00080 & 141 & 10.12 & 110K & 1200 & 120K \\
Oxford Radar RobotCar & 1.50 & 0.00058 & 161 & 11.52 & 95K & 1100 & 100K \\
D-Behavior (D-Set) & 1.78 & 0.00061 & 159 & 11.38 & 100K & 1100 & 100K \\
nuScenes by Motional & 2.38 & 0.00070 & 143 & 10.38 & 105K & 1200 & 120K \\
Comma2k19 & 1.98 & 0.00065 & 149 & 11.38 & 100K & 1200 & 120K \\
\bottomrule
\end{tabular}
\end{table*}

\autoref{table2} presents an in-depth analysis of the performance metrics of the Loihi-2 neuromorphic computing platform across various datasets. This table methodically details the power consumption (in watts), inference time (in seconds), throughput (in Giga Operations per second), energy efficiency (in picojoules per operation), computational complexity, and the neural network's capacity in terms of neurons and synapses for each dataset. The datasets included, such as AIODrive, Oxford Radar RobotCar, D-Behavior (D-Set), nuScenes by Motional, and Comma2k19, cover a range of applications from autonomous driving to real-world driving scenarios. This comparison highlights the adaptability and efficiency of the Loihi-2 in processing diverse and complex datasets. It underscores the chip's capability in high-speed processing while maintaining energy efficiency, which is crucial in advanced AI and machine learning applications.

\begin{figure}
    \centering
    \includegraphics[width = 0.4\textwidth]{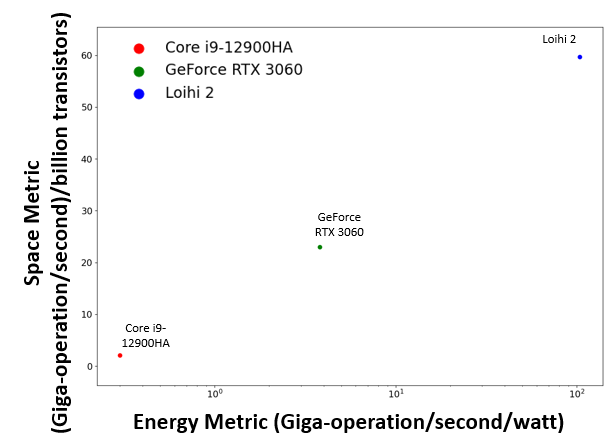}
    \caption{Comparison of energy efficiency, measured as giga-operation per second per billion transistors, against the energy required for one operation (giga-operation per second per watt) across various computing hardware. This plot illustrates the efficiency trade-offs between CPUs, GPUs, and Loihi-2 in processing.}
    \label{Fig:Energy_comsumptiom}
\end{figure}

\begin{table*}
\renewcommand{\arraystretch}{1.5}
\setlength{\tabcolsep}{10pt}
\centering
\caption{Comparisons with Previous Implementations on Various Datasets (Inference Times in ms)}
\label{table3}
\resizebox{\linewidth}{!}{
\begin{tabular}{|c|c|c|c|c|c|c|c|c|c|c|c|}
\hline
\textbf{Dataset} & \multicolumn{11}{c|}{\textbf{Inference Time (ms)}} \\
\hline
& Lopez et al. & Burnett et al. & Stacker et al. & RVF-Net & CRF-Net & RVNet & CRAFT & Liu et al. & Chen et al. & Echterhoff et al. & \textbf{Ours} \\
\hline
Hardware & FPGA & V100 GPU \& Xeon CPU & RTX 2080 GPU & Titan X & Titan XP & GeForce 1080 & RTX 3090 & RTX 2080Ti & GTX 1080 & RTX A6000 & \textbf{Loihi-2} \\
\hline
Oxford Radar RobotCar & - & 70 & - & - & - & - & - & - & - & - & \textbf{0.58} \\
\hline
D-Behavior (D-Set) & - & - & - & - & - & - & - & 330 & - & - & \textbf{0.61} \\
\hline
nuScenes by Motional & 18 & 700 & 36.7 & 44 & 43 & 17 & 244 & - & - & 50 & \textbf{0.70} \\
\hline
Comma2k19 & - & - & - & - & - & - & - & - & 0.01 & 4.1 & \textbf{0.65} \\
\hline
\end{tabular}}
\end{table*}

\autoref{table3} compares various implementations in terms of their inference times on different datasets, specifically focusing on advanced technologies in autonomous driving and image processing. The implementations by researchers are contrasted with our Loihi-2 implementation. The comparison underscores the performance capabilities of these diverse computational platforms in handling complex data processing tasks in real-world scenarios.

\section{Conclusion}
The conclusion drawn from our study indicates that the Loihi-2 neuromorphic chip excels in sensor data integration tasks. Our extensive evaluations have demonstrated that the chip performs these tasks faster and more efficiently than conventional computing methods. Notably, its low power consumption stands out, positioning it as an ideal candidate for high-demand computational applications where energy efficiency is paramount. The Loihi-2 chip's ability to rapidly and effectively process complex data streams suggests it could play a pivotal role in the evolution of autonomous systems, potentially leading to more advanced and sustainable technological solutions.

\section{Future Work}

This paper mainly uses the data generated for testing purposes to compare the chips in an ideal environment. In future work, we plan to use a real-time sensor fusion task to illustrate the practical results. Using real-time data will be enlightening while comparing the performance of Loihi-2 and others in the long run. Various domains, like defense, health, and the automotive industry, focus on improving their sensing capability using optical sensing and radar imaging \cite{yeong2021sensor,Fayyad2021sensors}. For example, new-generation autonomous vehicles combine Vision Cameras, Lidar, and Radar sensors to increase the sensing performance for safety issues, and processing power has a critical role in the sensor fusion \cite{yeong2021sensor}. In future work, we plan to continue with the industry to use Loihi-2 in their sensor fusion tasks, which will also be useful for real-time data generation.

\section{Acknowledgment}
We acknowledge the Temsa Research R\&D Center for their generous financial support and the reviewers for their invaluable insights and suggestions that significantly contributed to the enhancement of our paper.

\bibliographystyle{IEEEtran}
\bibliography{external}

\end{document}